\documentclass[aps,pre,twocolumn,amsmath,amssymb,superscriptaddress]{revtex4-1}
\usepackage[colorlinks,linkcolor=blue,citecolor=blue,urlcolor=blue]{hyperref}
\usepackage{graphicx}
\usepackage{txfonts}
\usepackage{dcolumn}

\begin{document}

\title{Multistage onset of epidemics in heterogeneous networks}
\author{Chao-Ran Cai}
%\email{ccr@nwu.edu.cn}%
\address{School of Physics, Northwest University, Xi'an 710069, China}
\address{Shaanxi Key Laboratory for Theoretical Physics Frontiers, Xi'an 710069, China}
\author{Zhi-Xi Wu}\email{wuzhx@lzu.edu.cn}
\affiliation{Institute of Computational Physics and Complex Systems, Lanzhou University, Lanzhou, Gansu 730000, China}
\author{Petter Holme}\email{holme@cns.pi.titech.ac.jp}
\affiliation{Tokyo Tech World Research Hub Initiative (WRHI), Institute of Innovative Research, Tokyo Institute of Technology, Nagatsuta-cho 4259, Midori-ku, Yokohama, Kanagawa, 226-8503, Japan}

\begin{abstract}
We develop a theory for the susceptible-infected-susceptible (SIS) epidemic model on networks that incorporate both network structure and dynamic correlations. This theory can account for the multistage onset of the epidemic phase in scale-free networks. This phenomenon is characterized by multiple peaks in the susceptibility as a function of the infection rate. It can be explained by that, even under the global epidemic threshold, a hub can sustain the epidemics for an extended period. Moreover, our approach improves theoretical calculations of prevalence close to the threshold in heterogeneous networks and also can predict the average risk of infection for neighbors of nodes with different degree and state on uncorrelated static networks.
\end{abstract}

\maketitle

\section{Introduction}
One can describe many systems in nature and society as a networked infrastructure for some spreading phenomena~\cite{newmanbook,Barrat}. Disease, information, economic shocks, rumors, opinions all spread over networks. Such situations are often described by models of epidemic spreading, even when the spreading is not concerning a disease. One of the canonical epidemic models is the susceptible-infected-susceptible (SIS) model. Figuratively, it describes an infection that leaves the individuals susceptible upon recovery. However, authors have used it for many other purposes---perhaps most prominently, it is the stochastic counterpart of Verhulst's logistic growth model of populations~\cite{nasell}. Understanding the SIS model is thus of broad scientific importance.

In this paper, we assume the SIS model is confined to a network of $N$ nodes~\cite{kiss2017mathematics}. We denote the number of infected nodes by $I$, and susceptible by $S=N-I$. Transmissions happen between network neighbors at a rate $r$. Simultaneously, infected nodes become re-susceptible at a rate $g$. 
Such dynamics exhibit a phase transition in which there is a critical infection rate (called epidemic threshold) that separates a disease-free (absorbing) state from an active stationary state (where a fraction of the population is infected). Theoretical research about this model focuses on predicting average prevalence (average number of infected nodes in the active state), calculating epidemic threshold, and times to extinction (time to get to an absorbing state in low infection rates)~\cite{ss_2,kiss2017mathematics,nasell,PhysRevLett.109.188701,PhysRevLett.117.028302,refId0,Holme_2018}.

Assuming that edges are placed independently at random according to some degree distribution (so small probability of short cycles) one can derive the prevalence analytically by a so-called  \textit{heterogeneous mean-field theory} (HMF)~\cite{hmf_1,hmf_2}. The epidemic threshold (in terms of the effective infection rate $\lambda=r/g$) in such a situation is $\lambda_c^{\mathrm{HMF}}=\langle k\rangle/\langle k^2\rangle$ where $\langle k\rangle$ and $\langle k^2\rangle$ are the first and second moment of the degree distribution, respectively.
Going beyond degrees, one can take the actual connectivity of the network into account through its adjacency matrix. 
A first step to capturing network structure effects is to apply a \textit{quenched mean-field theory} (QMF)~\cite{qmf_1,qmf_2}. That gives the thresholds $\lambda_c^{\mathrm{QMF}}=1/\Lambda$~\cite{qmf_1}, where $\Lambda$ is the largest eigenvalue of the adjacency matrix. However, also QMF fails to reproduce simulated results (see Appendix~\ref{App:A}), partly because it neglects dynamic correlations (for example, that the neighbor of an infected node has a higher chance of being infected than the average node).

Recently, there have been many attempts to take dynamic correlations into account in analytical theories.
These include the \textit{pair approximation}~\cite{pa_1,pa_2}, \textit{three-point approximation}~\cite{Ferreira2013}, \textit{effective degree} approach (ED)~\cite{ed_1,Gleeson_1,ed_2}, \textit{dynamic correlation} (DC)~\cite{dc}, and \textit{heterogeneous pair-approximation} (PHMF)~\cite{Mata_2014}. In particular, DC, which combines HMF and ED's advantages, could give accurate and straightforward analytical solutions to predict epidemic prevalence on the degree-uncorrelated static networks. However, these approaches only reveal the role of dynamic correlation and correctly predict the prevalence only when the infection rate is far above the epidemic threshold.

To obtain a theory that holds close to the threshold, one needs to build master equations that account for the network structure and dynamic correlations. This is the premise for the pair quenched mean-field (PQMF) approach~\cite{Mata_2013} and epidemic link equations~\cite{Matamalaseaau4212}. 
We know that any master equation-based approach (including PQMF) uses time iteration to accumulate the effects of long-distance correlations to obtain epidemic prevalence. That is to say, PQMF is a phenomenological theory, from which, however, we can't figure out explicitly how the network structure and the dynamic correlation are coupled. The answer of the question is just implicitly included in the set of master equations.

Besides the failures to predict the prevalence, a peculiar phenomenon often occurs in quasi-stationary state simulations on finite networks~\cite{PhysRevE.71.016129}. There commonly two or more peaks in the susceptibility curves $\chi=N (\langle \rho^2\rangle-\langle \rho\rangle^2)/\langle\rho\rangle$ appear for scale-free networks with $\gamma>3$. ($\chi$ measures fluctuations in the density of infected nodes and is commonly used to determine the critical infection rate in finite networks. $\rho$ is the fraction of infected nodes.)

This is at odds with $\chi$'s unique peak of in other types of interaction networks~\cite{th}, including random regular networks, Erd\H{o}s-R\'enyi networks, and scale-free networks with $\gamma<3$. 
The authors of Ref.~\cite{th,Castellano2012} speculated that the peak at small infection rate corresponds to the prediction by the QMF formula, and the peak at large infection rate corresponds to the $k$-cores collectively becoming active~\cite{Castellano2012}. (Active meaning that they can sustain the infection within themselves.) 
However, Mata and Ferreira showed the possibility of three or more susceptibility peaks~\cite{PhysRevE.91.012816}.
The authors of Ref.~\cite{PhysRevE.91.012816, PhysRevE.94.042308} argued that the multiple peaks are associated with the large gaps in the degree distribution among the nodes of highest degrees. 

In addition to the threshold on the simulations,
there has been a lot of theoretical work devoted to the understanding of epidemic threshold~\cite{chatterjee2009,mountford2013,PhysRevLett.105.218701,PhysRevLett.109.128702,PhysRevLett.111.068701,PhysRevE.87.062812,PhysRevResearch.1.033024,PhysRevX.10.011070} and the critical behavior around it~\cite{PhysRevE.88.032109,PhysRevE.93.032322}. Specifically, the authors of Ref.~\cite{chatterjee2009,mountford2013} rigorously proved that the epidemic threshold vanishes on random networks with power-law degree distributions.
Goltsev \textit{et al.}\ showed that the disease is mainly localized on a finite number of individuals when the effective infection rate $\lambda$ is slightly larger than the threshold $\lambda_c^{\mathrm{QMF}}$ on highly heterogeneous networks~\cite{PhysRevLett.109.128702}.
Mata and Ferreira proposed the PQMF to predict the first localized peak of the susceptibility curves~\cite{Mata_2013}. And Castellano \textit{et al.}\ gave an analytic solution~\cite{PhysRevX.10.011070} for the global threshold and showed that the threshold vanishes more slowly than predicted by the QMF theory. However, there is still no effective way to predict other localized peak.

In this paper, we develop a theory to analyse the coupling effect between the network structure and dynamic correlations directly, which works for the whole range of infection rate, leading to either population-wide outbreaks or small localized outbreaks on scale-free networks. Our theory can predict a more accurate prevalence both for population-wide outbreaks and localized spreading, and can predict the first and other localized peaks of the susceptibility curves.

\section{Model}
\subsection{Uncorrelated configuration model}
These scale-free static networks in this paper can be generated using an algorithm proposed by Catanzaro \textit{et al.}~\cite{catanzaro2005generation}, which is called the uncorrelated configuration model.
For each node we first assign it a degree $k_i$ according to the prescribed degree distribution, whose value is restricted to the range of [$k_{\min}$, $\sqrt{N}$], where $k_{\min}$ is the minimum degree of the node in the network and $N$ is the network size. And then we create a set of $k_i$ stubs that represent each of these edges with only a single tail connected to a node. If there is an uneven number of stubs, a random individual is given an extra stub. Finally, we construct the network by connecting pairs of these stubs chosen uniformly at random to make complete edges respecting the preassigned degrees. Self connections or duplicate edges between nodes are not allowed in the generation process. We note that the are several subtleties when implementing the configuration model by random stub-matching~\cite{fosdick}, and the method we use might introduce a small bias. This might explain some of the discrepancies between analytical and numerical results but should not invalidate the approach.

\subsection{SIS simulation procedure on static networks}
The SIS dynamic process in static networks can be simulated as follows~\cite{th,cota2017optimized}: First, we build a list $\nu$ for infected nodes and calculate the total transition rate $\omega=\sum_{i\in I}(g+rk_i)$ at the initial state, where $k_i$ is the degree of node $i$. The list $\nu$ and the total transition rate $\omega$ are constantly updated in one simulation, and the time is incremented by $dt=1/\omega(t)$. And then, there are two possible events happening at time $t+dt$: (1) With probability $gI/\omega$, an infected node $i$ is chosen with equal probability from $\nu$ and recovered. (2) With probability $\sum_{i\in I}(rk_i)/\omega$, an infected node $i$ is chosen from $\nu$ at random and accepted with probability $k_i/k_{\max}$, which is repeated until one choice is accepted. Finally, a neighbor of $i$ is chosen randomly. 
If the neighbor is infected, nothing happens; otherwise it becomes infected. We iterate the whole process longer than it typically takes for the system to reach the steady-state.

\subsection{The procedure of quasistationary simulation}
The simulations in this paper were performed using the quasi-stationary (QS) method~\cite{th}, in which every time the system tries to visit the absorbing state it jumps to one of the pre-stored active configurations. These pre-stored active configurations are updated constantly, i.e. , a pre-stored configuration is chosen randomly and replaced by the present active configuration with a probability $p_rdt$. After a relaxation time $t_r$, the QS probability $\overline{P}(n)$ that the system has $n$ infected individuals is computed during an averaging time $t_a$. 
From the distribution $\overline{P}(n)$, the moments of the activity distribution can be computed as $\left\langle\rho^k\right\rangle=\sum_n(n/N)^k\overline{P}(n)$.
And then, the epidemic threshold can be obtained by the maximum value of the susceptibility $\chi$, whose value is defined as $\chi=N\left\langle\rho^2\right\rangle-\left\langle\rho\right\rangle^2)/\left\langle\rho\right\rangle$. In this paper, the number of active configurations is set as 100, $p_r=0.02$, $t_a=3t_r$, and $t_r$ depends on $N$ and $\lambda$.

\section{Theory and results}
\subsection{Population-wide outbreaks}
We denote $p_k$ and $q_k$, respectively, as the probabilities of reaching an arbitrary infected individual by following a randomly chosen edge from a susceptible and infected individual of degree $k$~\cite{dc}. 
To calculate the $p_k$ and $q_k$ on scale-free networks, our analysis takes five steps.

First, we choose a node $i$ whose degree is $k$, which is the only known condition. And then, we divide the whole scale-free network into two subnetworks. One is a star subnetwork (denoted \textit{star}) where the center node is node $i$. The other is the scale-free network minus the star subnetwork (denoted \textit{minus-star}). Note that we do not ignore triangles. For any triangle $[ijj']$ where the nodes $j$ and $j'$ are the neighbors of $i$, the edges $[ij]$ and $[ij']$ belong to the former subnetwork, and the edge $[jj']$ belongs to the latter subnetwork.

Second, from the \textit{star}, we can get the probability $A$ that any node $j$ (a neighbor of $i$) is infectious. However, the calculation of this probability currently does not consider the node $j$'s neighbors except node $i$.

Third, for a large $k$, node $i$ has many neighbors. For a small $k$, there are many nodes with degree $k$ on a scale-free network. That is, there are many nodes $j$ regardless of $k$. For the \textit{minus-star}, we can, by mean-field theory, get the average probability $B(k')$ that any node $j''$ with degree $k'$ is infectious. We can get the joint probability $C$ that the node $j''$ links the node $i$ and the node $j''$ is infected.

Fourth, we do not need to calculate these exact results $A$ and $C$, because they will affect each other on the whole scale-free network. Here, we assume approximately $C$ converges to $x$. And then, we preset that a fraction $x$ of the nodes are infected nodes all the time (they are fixed in the infection state) on a star network where the degree of its center node is $k$.

Fifth, $p_k$ can be calculated on the star network with permanently infected nodes (that we mentioned in Step 4), but it still contains an unknown $x$. We solve the $x$ by using the other global relation, i.e.\ Eq.~\eqref{e.13}. As a result, the probability $p_k$ (or $q_k$) can be divided into two parts: one from the nearest-neighbor network structure of the chosen node itself; the other representing the dynamic correlation coming from the whole network except the chosen node. The latter can be approximated as a fixed value on uncorrelated networks. Suppose all neighbors of one center node are identical, we consider a star subnetwork with a fraction $x$ of permanently infected nodes to calculate the probabilities $p_k$ and $q_k$, where $k$ is the degree of the star. Note that $x$ represents a joint probability, which cannot be replaced by the prevalence $\rho$ of scale-free networks.

\begin{figure}
\includegraphics[width=\columnwidth]{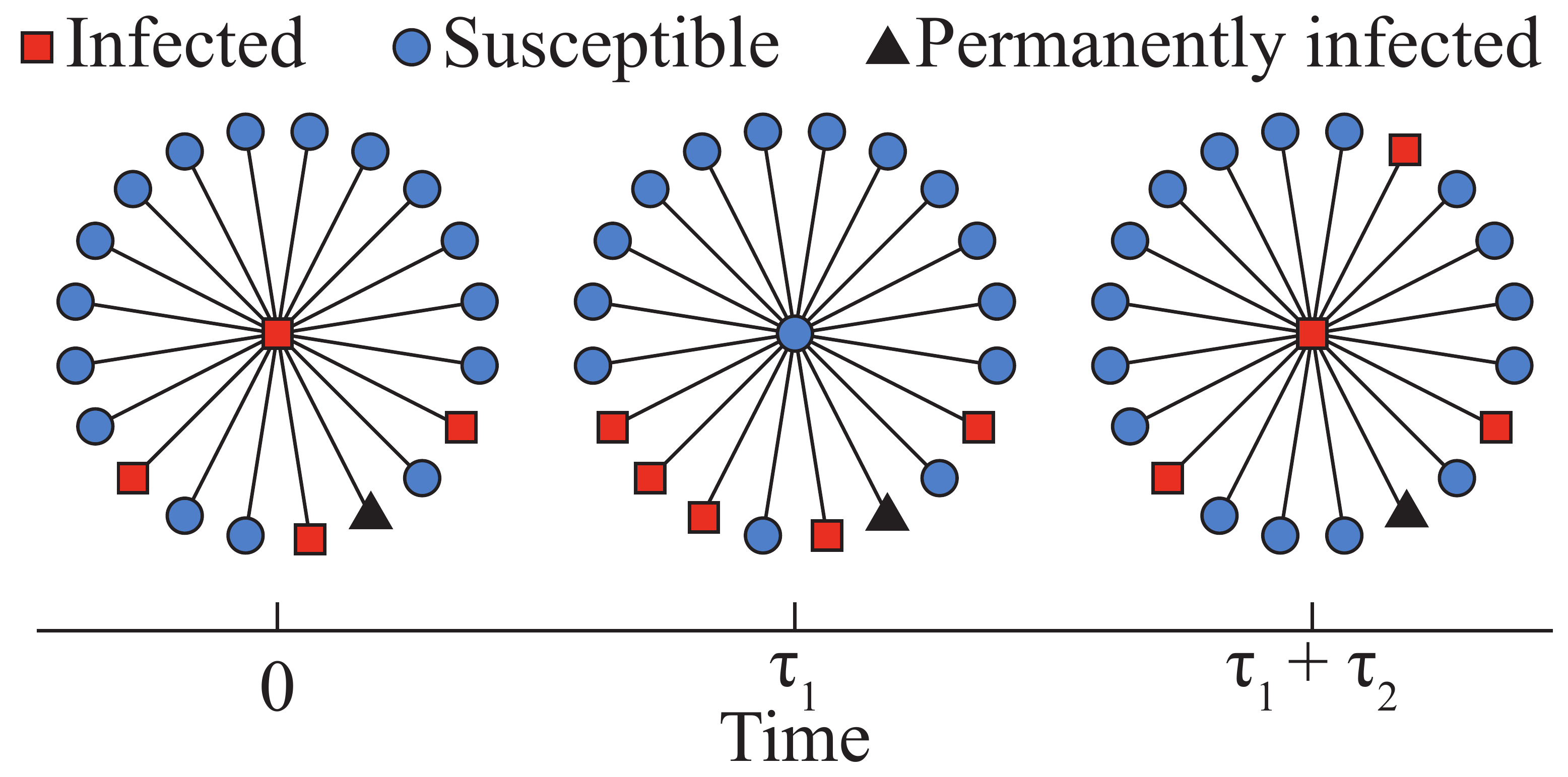}
\caption{Schematic illustration of one state cycle of the center node on a star subnetwork with a nonzero fraction of permanently infected leaf nodes.}\label{fig.1}
\end{figure}

In Fig.~\ref{fig.1}, we show a schematic illustration of one state cycle of the center node.
We denote the number of infected leaf nodes at time $t$ by $I(t)$. 
To make it easier for the reader to understand our model, we summarize our ideas with simple words before the formula derivation begins:
First, we derive the relationship between $I(t)$ and the joint probability $x$ on a star subnetwork with nonzero permanently infected leaf nodes at any given time (see Eqs.~\eqref{e.1}--\eqref{e.3}); Second, according to the definition of $p_k$ (i.e. the relationship between $I(t)$ and $p_k$), we can estimate the relationship between $p_k$ and $x$ (see Eqs.~\eqref{e.4}--\eqref{e.11}); Third, we solve $p_k$ and $x$ by introducing another global law, Eq.~\eqref{e.14}.

Since both $I(0)$ and $I(\tau_1)$ have peaked distributions (see Appendix~\ref{App:B}), we use their expected values as approximations. From Fig.~\ref{fig.1}, we assume that the center node is infected at time $0$, and there are $n$ infected leaf nodes, \textit{i.e.}, $I(0)=n$. After a time interval $\tau_1$, the center node is cured but the number of infected leaf nodes rises to $m$ (so $I(\tau_1)=m$). Then, $I(t)$ in the recovery process is determined by
\begin{equation}\label{e.1}
I(t)=I(0)+\int_0^t[k-I(t')]rdt'-\int_0^t[I(t')-xk]gdt',
\end{equation}
where the second and third terms on the right hand side are the number of newly infected, and recovered, leaf nodes in the time interval $[0,t]$, respectively. After a time $\tau_2$, the center node becomes reinfected, and the number of infected leaf nodes reduces to $I(\tau_1+\tau_2)=n$. $I(t)$ in the infection process is
\begin{equation}\label{e.2}
I(t)=I(\tau_1)-\int_{\tau_1}^{t}g[I(t')-xk]dt',
\end{equation}
where the second term on the right hand side is the number of newly recovery leaf nodes for times in $[\tau_1,\tau_1+t]$.
Combining Eqs.~\eqref{e.1} and~\eqref{e.2} gives
\begin{equation}\label{e.3}
I(t)=\left\{
   \begin{array}{cl}
   \frac{kr+xkg-e^{-(g+r)t}\left[kr+xkg-(g+r)n\right]}{g+r}, & 0\leq t \leq\tau_1;\\
   xk+(m-xk)e^{-g(t-\tau_1)}, & \tau_1\leq t\leq\tau_2.\\
   \end{array}
  \right.
\end{equation}

We define $P(\tau_1)$ as the probability density of the duration $\tau_1$ in the recovery process.
Since the total recovery rate is $g\tau_1$ during the time interval $[0,\tau_1]$, the center node's recovery probability can be calculated as $1-e^{-g\tau_1}$ in this interval~\cite{PhysRevE.93.032314}. Then, we obtain $P(\tau_1)=ge^{-g\tau_1}$, by the definition of cumulative distributions. We define $P(\tau_2)$ as the probability density of the duration $\tau_2$ in the infection process. We can calculate the infection probability of the center node as $1-\exp^{-rI(t)dt}$ in a time interval $[t,t+dt]$. The cumulative infection probability of time interval $[\tau_1,\tau_1+\tau_2]$ can be calculated as $1-\prod_{t=\tau_1}^{\tau_1+\tau_2}\exp^{-rI(t)dt}$. Combining this with Eq.~\eqref{e.3}, we can obtain the probability density $P(\tau_2)$,
\begin{align}\label{e.4}
P(\tau_2)=&r[xk+(m-xk)e^{-g\tau_2}]  \nonumber\\
&\times \exp\left[-r\left(xk\tau_2+\frac{m-xk}{g}(1-e^{-g\tau_2})\right)\right].
\end{align}
Moreover, combining Eq.~\eqref{e.3} and Eq.~\eqref{e.4}, we can further find the relation below,
\begin{equation}\label{e.5}
\int P(\tau_2)d\tau_2\int_{\tau_1}^{\tau_1+\tau_2}I(t')rdt'=1.
\end{equation}
Eq.~\eqref{e.5} applies to any star subnetwork with non-zero permanently infected leaf nodes, which implies that the disease in Fig.~\ref{fig.1} will never become extinct.

Next, we average both sides of Eq.~\eqref{e.2} by $\tau_2$ and set $t=\tau_1+\tau_2$.
Considering Eq.~\eqref{e.5}, the expected decrease of $I(t)$ is $g/r-xkg\langle\tau_2\rangle$ in the infection process. Thus, we have
\begin{equation}\label{e.6}
n=m-\left(\frac{g}{r}-xkg\langle\tau_2\rangle\right),
\end{equation}
where $\langle\tau_2\rangle=\int \tau_2P(\tau_2)d\tau_2$ is the average duration of infection.
Combining Eq.~\eqref{e.3} with Eq.~\eqref{e.6} gives the number of infected leaf nodes at time $t=\tau_1$
\begin{equation}\label{e.7}
m=\frac{kr+xkg}{g+r}-\frac{g}{r}\frac{g}{g+r}+xkg\langle\tau_2\rangle\frac{g}{g+r},
\end{equation}
Finally, according to the definition of $p_k$ and $q_k$, 
\begin{subequations}\label{e.8}
\begin{align}
    p_k&=\frac{\int\left[\int_{\tau_1}^{\tau_1+\tau_2} I(t)dt\right]P(\tau_2)d\tau_2}{\int k\tau_2P(\tau_2)d\tau_2}, \\
    q_k&=\frac{\int\left[\int_0^{\tau_1} I(t)dt\right]P(\tau_1)d\tau_1}{\int k\tau_1P(\tau_1)d\tau_1},
    \end{align}
\end{subequations}
we can get
\begin{subequations}\label{e.9}
\begin{align}
    \label{e.9a}
    p_k&=x+\frac{kr-xkr-g^2/r+xkg^2\langle\tau_2\rangle}{g(g+r)k} \frac{\langle1-e^{-g\tau_2}\rangle}{\langle\tau_2\rangle}, \\
   \label{e.9b}
    q_k&=\frac{r+xg}{g+r}-\frac{g}{k(g+r)}\left(\frac{g}{r}-xkg\langle\tau_2\rangle\right),
    \end{align}
\end{subequations}
where $\langle1-e^{-g\tau_2}\rangle=\int(1-e^{-g\tau_2})P(\tau_2)d\tau_2$.
Here, we are able to rewrite $p_k$ to a simpler formula $1/(rk\langle\tau_2\rangle)$ by using the Eq.~\eqref{e.5}, but this formula is sensitive to the upper limit of the integral, where the upper limit is set as $\tau_2=50$ in this paper. 

Moreover, we can obtain the following inequality between $p_k$ and $q_k$ by combining Eq.~\eqref{e.7} and Eq.~\eqref{e.9},
\begin{equation}\label{e.10}
q_k-p_k=\left(\frac{m}{k}-x\right)\left(1-\frac{1}{g} \frac{\langle1-e^{-g\tau_2}\rangle}{\langle\tau_2\rangle}\right)>0.
\end{equation}
Ref.~\cite{dc} also presents this relation, but without a derivation from an underlying theory.

As we are modeling a stochastic process, there might be no infected leaf nodes other than the permanently infected ones. This situation becomes typical for small $k$ or small $\lambda$, which means that $I(t)< xk$. When this situation occurs, Eq.~\eqref{e.6} becomes inaccurate because of the average loss of $I(t)$ might not be $g/r-xkg\langle\tau_2\rangle$.
We can circumvent this problem by using the lower boundary $I(0)=xk$ for small $k$ or small $\lambda$.
In particular, by replacing the Eqs.\ \eqref{e.6}--\eqref{e.7} with $n=xk$, we obtain new estimates for $p_k$ and $q_k$:
\begin{subequations}\label{e.11}
\begin{align}\label{e.11a}
p_k&=x+\frac{r(1-x)}{g(2g+r)}\frac{\langle1-e^{-g\tau_2}\rangle}{\langle\tau_2\rangle},\ \ x>0
\\ \label{e.11b}
q_k&=1-\frac{2g(1-x)}{2g+r},\ \ x>0.
\end{align}
\end{subequations}
Here, Eq.~\eqref{e.11} is not appropriate when $x=0$, because of the high chance of extinction. 

Now, let us turn to the SIS dynamics on scale-free networks. When a scale-free network is in its epidemic state, we assume, for theoretical purposes, that there is a fixed fraction $x>0$ of infected nodes. As long as the epidemic process is in its quasi-stationary~\cite{nasell} state, the total recovery rate of $I_k$ must be equal to the total infection rate of $S_k$~\cite{hmf_1}. That means
\begin{equation}\label{e.12}
gI_k=rkp_k (N_k-I_k),
\end{equation}
where $N_k=S_k+I_k$ and $S_k$ ($I_k$) is the number of susceptible (infected) individuals with degree $k$.
In the framework of DC~\cite{dc}, we still require that the expected variations in the total recovery rate and total infection rate are equal in the steady state:
\begin{equation}\label{e.13}
\sum_{jk}\frac{S_{k,j}jr-I_{k,j}g}{\omega}[r(k-j)-g]=0,
\end{equation}
where $\omega=\sum_k\sum_j(S_{k,j}jr+I_{k,j}g)$ is the total rate and $S_{k,j}$ ($I_{k,j}$) is the number of susceptible (infected) nodes which have $j$ infected nodes among the total $k$ neighbors. Following Ref.~\cite{dc}, this reduces to
\begin{equation}\label{e.14}
\sum_k\frac{rkp_kN_k}{g+rkp_k}\left[(k-1)(p_k-1)+\frac{g}{r}\right]=0.
\end{equation}

For a given pair of $r$ and $g$ on a scale-free network, numerical calculation is divided into two major steps. First, we get $p_k$ for a preset $x$ by using Eqs.~\eqref{e.4}--\eqref{e.11}. Then, combining $p_k$ with the Eq.~\eqref{e.14}, we confirm whether the preset $x$ is a solution. In particular, we are able to calculate the probability $p_k$ in the steady state as follows:
(i) Give any $x'\in(0, 1]$, the values of $m$ and $n$ are calculated by Eqs.~\eqref{e.4}--\eqref{e.7};
(ii) The values of various $p_k$ are calculated by Eq.~\eqref{e.9} or Eq.~\eqref{e.11};
(iii) Plugging the set of $p_k$ into Eq.~\eqref{e.14}, and the value $x$ is the solution when Eq.~\eqref{e.14} is valid.
(iv) Inserting the $x$ into Eq.~\eqref{e.9} and Eq.~\eqref{e.11} to solve $p_k$.

\begin{figure}
\includegraphics[width=0.9\columnwidth]{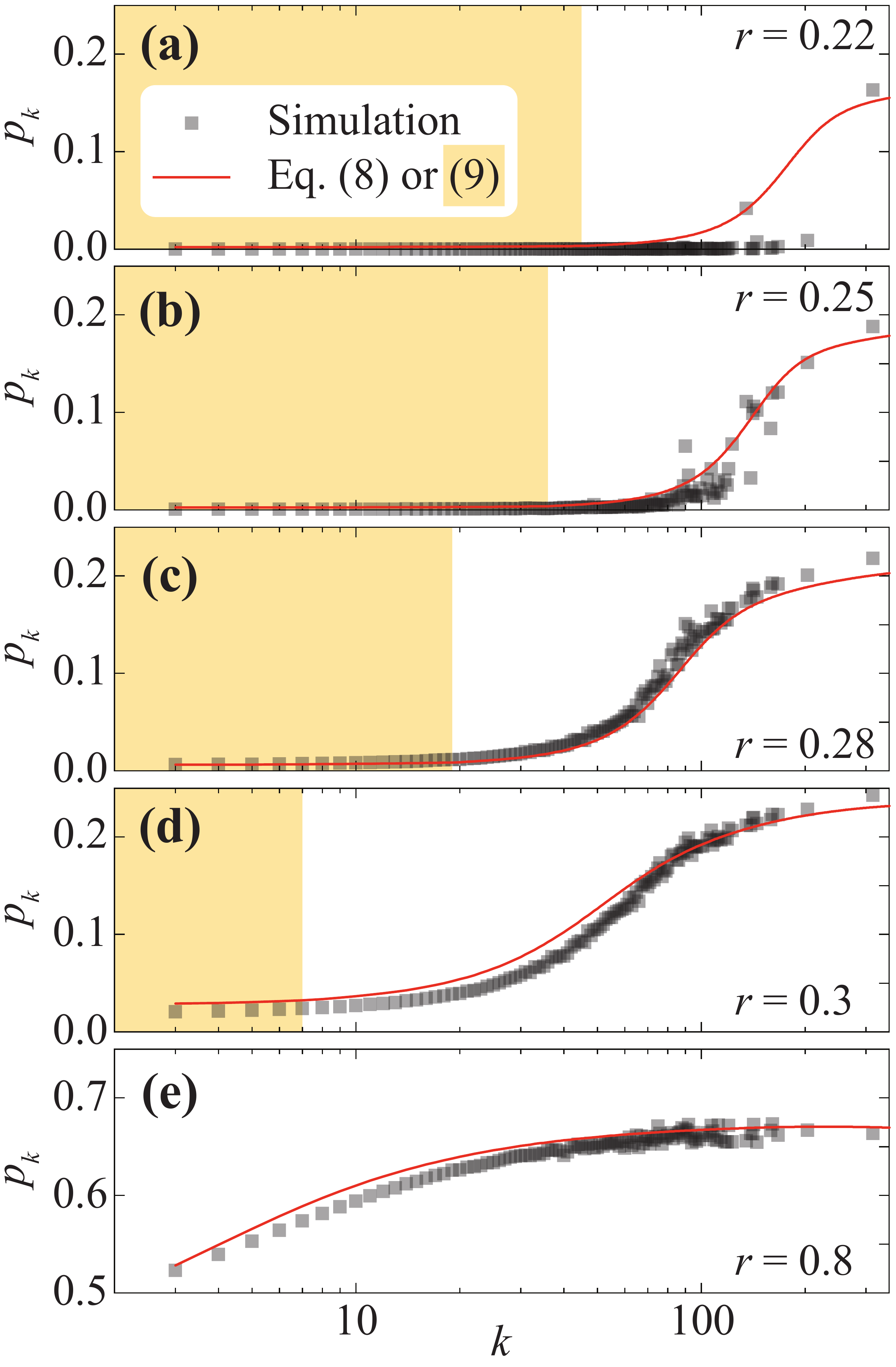}
\caption{$p_k$ as a function of degree $k$ on a scale-free network. These panels show simulations on one realization of uncorrelated configuration model with minimum degree $3$, $\gamma=4.5$ and $N=10^7$. Parameters: $g=1$, (a) $r=0.22$; (b) $r=0.25$; (c) $r=0.28$; (d) $r=0.3$; (e) $r=0.8$, the relaxation time $t_r=10^6$ in (a)--(d) and $t_r=5\times10^3$ in (e). The shaded areas in (a)--(d) indicate the position where Eq.~\eqref{e.11} is valid.}\label{fig.2}
\end{figure}

\begin{figure*}
\includegraphics[width=\textwidth]{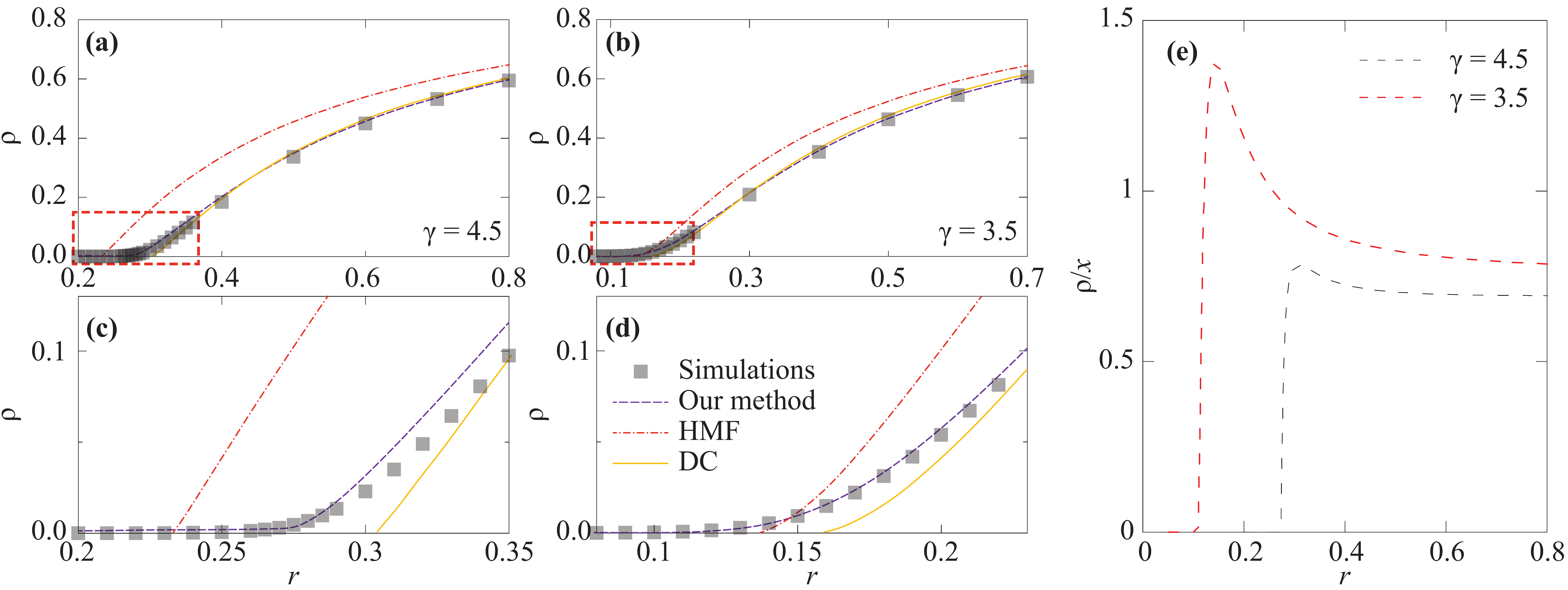}%\includegraphics[width=0.645\columnwidth]{figadd.pdf}
\caption{(a)--(d) The prevalence $\rho$ and (e) the ratio $\rho/x$ as a function of infection rate $r$ on uncorrelated static scale-free networks. Parameters are $g=1$, minimum degree $3$, $N=10^7$; $\gamma=4.5$ (a) and (c); $\gamma=3.5$ (b) and (d). Panels (c) and (d) are blow-ups of the boxed regions of panels (a) and (b), respectively. Simulation results are averaged over 20 independent runs. The values of panel (e) are from our theoretical results in (a) and (b).
}
\label{figrho}
\end{figure*}

In Fig.~\ref{fig.2}, we plot $p_k$ as a function of degree for several infection rates $r$ for the SIS model on a scale-free network with $\gamma=4.5$, $N=10^7$, and $k_{\max}=314$. As $r$ increases, the $k_{\max}$-star subnetwork (the star subnetwork of the largest-degree node) will become active before the other large-degree star networks.  
As $r$ increases, more star subnetworks will become active until the system reaches the global epidemic threshold ($x\to0$). 
We show that our theoretical method captures $p_k$'s properties beyond the global threshold in Fig.~\ref{fig.2}(c)--Fig.~\ref{fig.2}(e) and near it in Fig.~\ref{fig.2}(b). It only fails for the smallest infection rates, see Fig.~\ref{fig.2}(a).

For sufficiently low infection rates, the epidemic is localized around one or a few high-degree star subnetworks in Fig.~\ref{fig.2}(a).
Our theoretical method only captures two points, which implies that these two stars may be approximately continuously activated. By the actual SIS dynamics, the epidemics at any star network would go extinct after a average lifetime $T(k,r,g)$~\cite{PhysRevLett.111.068701,PhysRevX.10.011070}.
So, there are many long intervals on star subnetworks between deaths and reactivations, which leads to $p_k\to 0$.
Meanwhile, the mean-field treatment in Eqs.~\eqref{e.1}--\eqref{e.3} is no longer appropriate in localized spreading.
These contrast with when $r$ is larger, and many star subnetworks are active. Then the infected nodes are, as assumed by mean-field theory, more evenly distributed across the network.
This explanation also implies HMF, PHMF, and DC are not suitable for localized spreading. They mistakenly give a finite threshold on scale-free networks with $\gamma>3$, but their threshold expressions are still valuable on networks with no localized spreading, such as random regular networks, Erd\H{o}s-R\'enyi networks, and scale-free networks with $\gamma<3$. 

Finally, combining these $p_k$ with Eq.~\eqref{e.12}, we can obtain the more accurate prevalence $\rho$ in population-wide outbreaks ($x>0$), that is, 
\begin{equation}\label{e.15}
\rho=\frac{1}{N}\sum_kI_k=\sum_k\frac{\lambda kp_kP(k)}{1+\lambda kp_k},
\end{equation} 
where $\lambda=r/g$ is the effective infection rate and $P(k)$ is the degree distribution of the network. In Fig.~\ref{figrho}, we show numerical and approximate results of the SIS dynamics on a scale-free network with $N=10^7$ and $\gamma=4.5$ and $3.5$. We find that our theoretical approach's estimations of prevalence match those from stochastic simulations well on scale-free networks and are much better than the results predicted by the heterogeneous mean-field theory and the dynamic correlation method~\cite{hmf_1,dc}. Here, we only compare HMF and DC in Fig.~\ref{figrho} because they do not require the iterations of a large body of equations with respect to time (where PQMF, PHMF, ED does) to obtain the epidemic prevalence, which is the same as our theory. In Fig.~\ref{figrho}(e), we show the ratio between the epidemic prevalence $\rho$ and the joint probability $x$ as a function of infection rate $r$. Clearly, $x$ fails to approximate $\rho$ for low enough $r$. 

\subsection{Localized spreading}
\begin{figure*}
\includegraphics[width=0.8\textwidth]{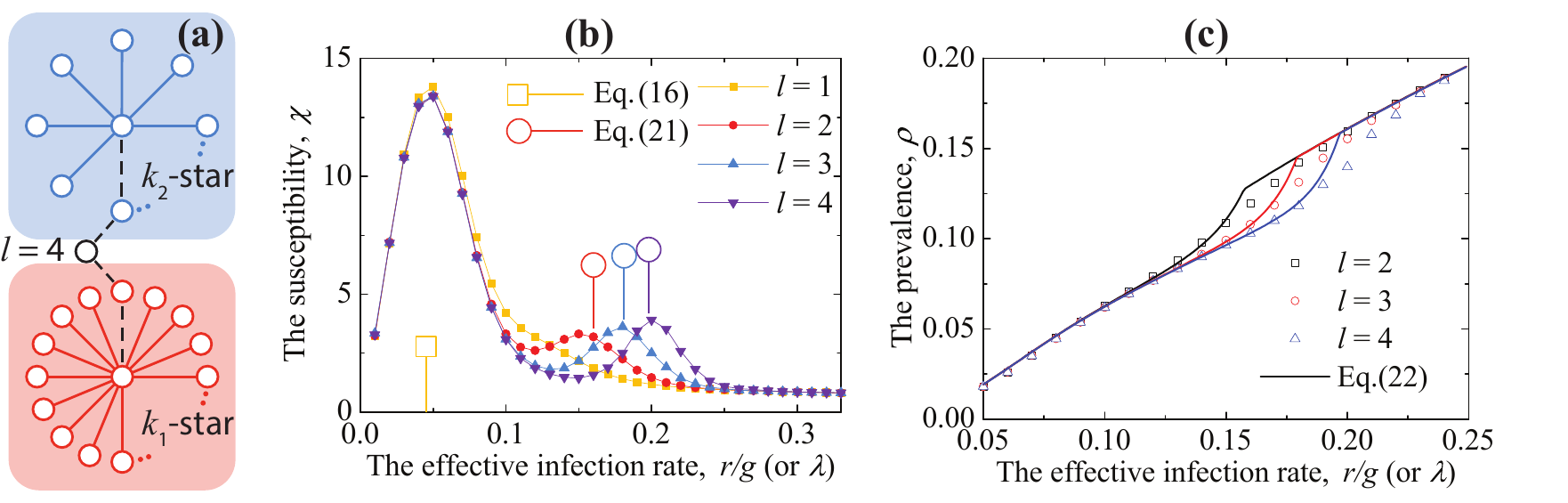}
\caption{Panel (a) illustrates a network consisting of two star subnetworks a distance $l$ apart. Dashed lines show the paths between hubs. Panels (b) and (c) show the susceptibility $\chi$ and prevalence $\rho$ as a function of the effective infection rate $\lambda$ for the network in (a). The parameters are $k_1=1000$ and $k_2=300$. 
Simulations on one realization with the relaxation time $t_r=10^6$.
}\label{fig.4}
\end{figure*}

The successive activation of star subnetworks with increasing $r$ is a natural explanation for the multi-peak phenomenon of the susceptibility $\chi$ observed in the literature~\cite{PhysRevE.91.012816,PhysRevE.93.032322}. If this scenario is correct, the separation of the hubs should affect the location of the peaks.
Comparing with the double random regular network of Ref.~\cite{PhysRevE.91.012816} which is formed by two random regular networks connected by a single edge, we consider a network consisting of two stars where the two center nodes are separated by $l$ edges in Fig.~\ref{fig.4}(a).

In order to deal with localized spreading, we use the local condition (separated distance $l$) instead of those global conditions in Eq.~\eqref{e.12} and Eq.~\eqref{e.14}. For $l=1$, the $x$ of $k_2$-star approximately equal to $1/k_2$ which is a finite value when the infection rate is slightly larger than the threshold of $k_1$-star. According to the Eq.~\eqref{e.5}, $k_2$-star is also active when $k_1$-star is active, which means that there is just one peak.
When we set $x=0$ in Eqs.~\eqref{e.6}--\eqref{e.7}, the threshold of $k_1$-star can be easily predicted as an isolated star network, that is
\begin{equation}\label{e.16}
\lambda_{c_1}(k)=\frac{1+\sqrt{1+8k}}{2k}.
\end{equation}
In Fig.~\ref{fig.4}(b), we can see that the Eq.~\eqref{e.16} captures the first peak very well.

\begin{figure}
\includegraphics[width=\columnwidth]{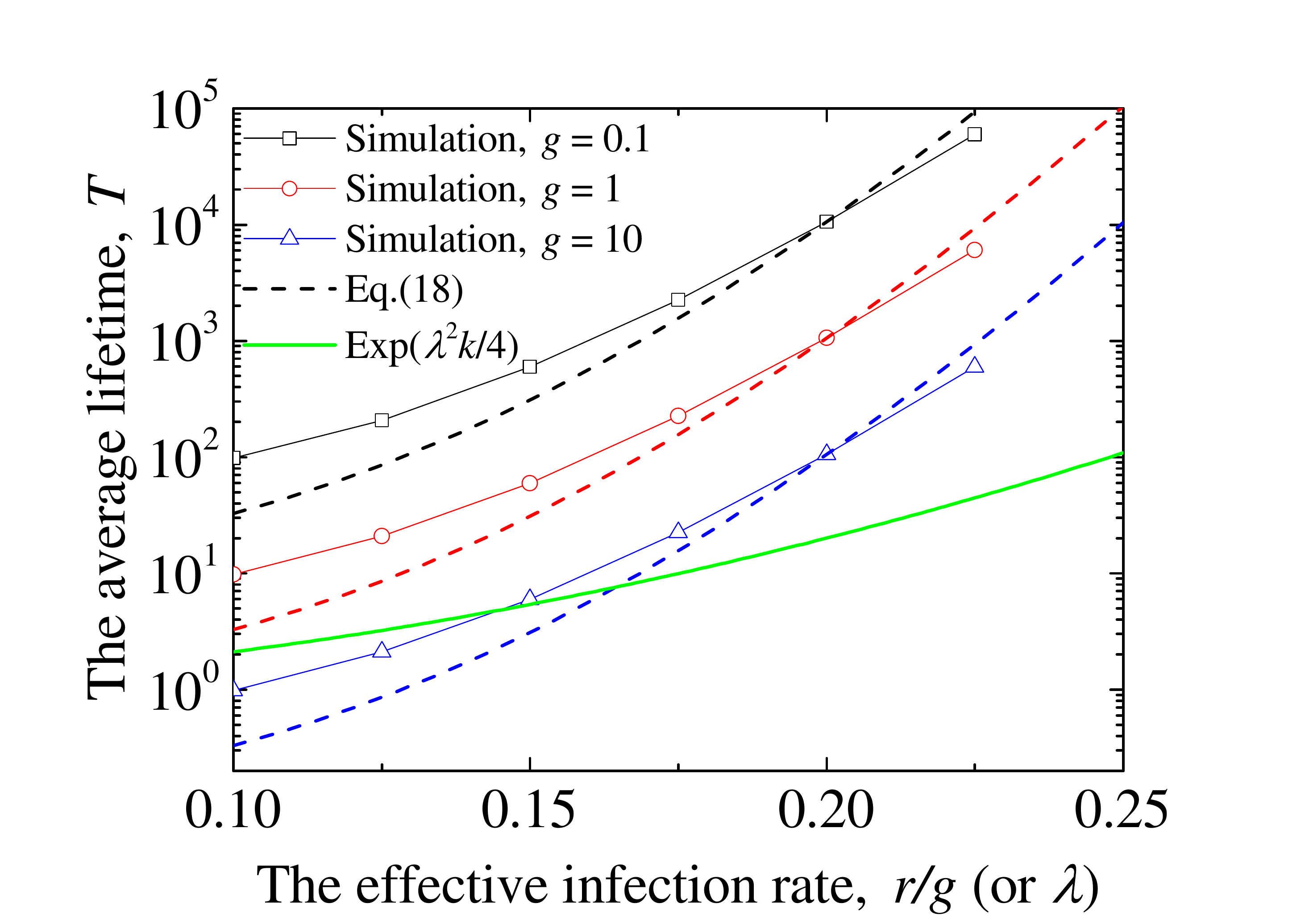}
\caption{The average lifetime $T$ as a function of the $\lambda$ on a $k$-star network. The degree of the center node is $k=300$. Simulation results are averaged over $5\times10^4$ independent runs.}\label{fig.5}
\end{figure}

For $l\geq2$, the $x$ of $k_2$-star is $\left(r/g\right)^{l-2}p_{k_1}/k_2$ which is very small and vanish for large $l$.
After setting $x=0$, the Eq.~\eqref{e.5} can be rewritten as
\begin{equation}\label{e.17}
\int P(\tau_2)d\tau_2\int_{\tau_1}^{\tau_1+\tau_2}I(t')rdt'=1-\exp^{-\frac{rm}{g}}\left(1+\frac{rm}{g}\right),
\end{equation}
where $P(\tau_2)=rm\exp^{-g\tau_2}\exp^{-\frac{rm}{g}(1-e^{-g\tau_2})}$. The Eq.~\eqref{e.17} means that the isolated star network will die with a probability $Y=\left(1+\frac{rm}{g}\right)\exp^{-\frac{rm}{g}}$. And then, its average lifetime $T(k,r,g)$ can be calculated as
\begin{align}\label{e.18}
T(k,r,g)&=\left[\langle\tau_1\rangle+\langle\tau_2(k)\rangle\right]\sum_n nY(1-Y)^{n-1}  \nonumber\\
&=\frac{1+g\langle\tau_2(k)\rangle}{g+rm(k)}\exp^{\frac{rm(k)}{g}},
\end{align}
where $\langle\tau_1\rangle=1/g$ and $m(k)=(kr^2-g^2)/[r(g+r)]$. In Fig.~\ref{fig.5}, we can see that the Eq.~\eqref{e.18} is a more accurate prediction of average lifetime on $k$-star network than that given from Ref.~\cite{PhysRevLett.111.068701,PhysRevX.10.011070}.
Note that $T(k,r,g)$ depends on the individual values of $r$ and $g$, not the ratio $r/g$. 

We denote $\eta(k)$ as the probability that any leaf node is in the infected state on an active $k$-star network, that is 
\begin{equation}\label{e.19}
\eta(k)=\frac{q_{k}\langle\tau_1\rangle+p_{k}\langle\tau_2(k)\rangle}{\langle\tau_1\rangle+\langle\tau_2(k)\rangle}.
\end{equation}
The upper boundary of Eq.~\eqref{e.19} is $\eta(k)<r/(g+r)$ by using the Eq.~\eqref{e.10}.
And then, we can calculate the probability of $k_2$-star being active per unit time on the two star network in Fig.~\ref{fig.4}(a),
\begin{equation}\label{e.20}
f(k_1,k_2,r,g)=\eta(k_1)  \left[r\left(\frac{r}{g}\right)^{l-2}\right]T(k_2,r,g).
\end{equation}
Note that the recovery process of the leaf nodes of $k_1$-star has been considered by $\eta(k_1)$.
The $f(k_1,k_2,r,g)\geq1$ means that $k_2$-star subnetwork can continuously be active.
Combining $f(k_1,k_2,r,g)=1$ and $\langle\tau_1\rangle\gg\langle\tau_2\rangle$ with Eqs.~\eqref{e.18}--\eqref{e.20}, we can predict the activation point of $k_2$-star subnetwork (the lower peak of $\chi$) as follows,
\begin{equation}\label{e.21}
\frac{\lambda_{c_2}^{l-1}m(k_1)}{k_1[1+\lambda_{c_2} m(k_2)]}\exp^{\lambda_{c_2} m(k_2)}=1.
\end{equation}
In Fig.~\ref{fig.4}(b), we can see that the activation point of the $k_2$-star subnetwork strongly depends on the distance $l$ between the center nodes, and the Eq.~\eqref{e.21} captures them very well.
As $l$ decreases, the lower peak moves toward small $r$ until extinction ($l=1$), which all can be predicted quantitatively by the localized analysis of our theory.

Finally, based on the localized analysis above, we can predict the localized prevalence as follows, 
\begin{equation}\label{e.22}
\rho=\frac{1}{N}\sum_{\Omega}\left\{\overline{I(k)}\ F(k)\ H[\lambda-\lambda_{c_1}(k)]\right\}
\end{equation}
where $\Omega$ contains all star subnetworks, $\overline{I(k)}=\frac{\langle\tau_1\rangle}{\langle\tau_1\rangle+\langle\tau_2(k)\rangle}+k\eta(k)$
is the average number of infected individuals when the $k$-star subnetwork is active, the $F(k)H[\lambda-\lambda_{c_1}(k)]$ represents the activation probability of the $k$-star subnetwork. The $F(k)$ is a ramp function whose value is $\min\{f(k_{\max},k,r,g),1\}$ and the $H[\lambda-\lambda_{c_1}(k)]$ is a Heaviside step function whose value is zero for negative argument and one for positive argument. In Fig.~\ref{fig.4}(c), we can see that the estimations in Eq.~\eqref{e.22} match those from stochastic simulations quite well.

\subsection{Epidemic thresholds}
Since epidemic thresholds are defined only in the $N\rightarrow\infty$ limit, the peaks of $\chi$ from our quasi-stationary simulations are, technically speaking, not true thresholds [$\lambda_{c_1}
(\text{$k_{\max}$})<\lambda<\lambda_c^{\text{global}}$].
For uncorrelated static scale-free networks $\lambda_{c_1}
(\text{$k_{\max}$})$ is related to $\lambda_c^{\text{global}}$~\cite{PhysRevX.10.011070} via
\begin{equation}\label{eq:localization}
\frac{\lambda_c^{\text{global}}}{\lambda_{c_1}
(\text{$k_{\max}$})} \sim
\frac{\ln(k_{\max})k_{\max}^{-1/2}}
{\sqrt{2}k_{\max}^{-1/2}}
\sim\frac{\ln\left(k_{\min}N^{1/(\gamma-1)}\right)}{\sqrt{2}},
\end{equation}
where we have used the natural degree cut-off~\cite{doro}, $k_{\mathrm{cut}}/k_{\min}\sim N^{1/(\gamma-1)}$ as an estimate of $k_{\max}$.
Equation~\eqref{eq:localization} means that $\lambda_{c_1}
(\text{$k_{\max}$})$ decreases faster than $\lambda_c^{\text{global}}$ with increasing $N$, which leaves a finite range of $\lambda$ where epidemic localization can emerge so that the outbreak survives in the $k_{\max}$-star subnetwork, but not in the rest of the network. We also note that is $\gamma>3$, hubs will be separated in large networks since the probability that nodes of degrees $k_1$ and $k_2$ are connected is proportional to $k_1k_2/N\langle k\rangle$ which goes zero with $N$.

For networks with more homogeneous degree distributions, such as random regular networks and Erd\H{o}s-R\'enyi networks, the largest degree is not large enough to ensure $\lambda_{c_1}
(\text{$k_{\max}$})<\lambda_c^{\mathrm{DC}}$, where $\lambda_c^{\mathrm{DC}}=\langle k\rangle/(\langle k^2\rangle-\langle k\rangle)$ is the epidemic threshold predicted by DC or PHMF~\cite{dc,Mata_2014}. Thus, a $k_{\max}$-star subnetwork would not become active until $\lambda$ exceeds the true threshold. Similarly, for the scale-free network with $2<\gamma<2.5$, we also have the relations of $\lambda_c^{\mathrm{QMF}}<\lambda_{c_1}
(\text{$k_{\max}$})$ and $\lambda_c^{\mathrm{DC}}<\lambda_{c_1}
(\text{$k_{\max}$})$ (see Appendix~\ref{App:C}). Thus, also for these networks, we expect one susceptibility peak, not because the largest degree is too small, but because the largest-degree nodes are connected.

Note that the discussion of the scale-free network with $2<\gamma<2.5$ mentioned above is from the uncorrelated configuration model. For a degree-degree disassortative correlations network, low degree nodes act as bridges linking the hubs, i.e. , the situation similar to Fig.~\ref{fig.4}(a) can easily arise. Interestingly, we may see the multi-peak phenomenon on the degree-degree disassortative correlations of the scale-free network with $2<\gamma<2.5$.

\begin{figure}
\includegraphics[width=\columnwidth]{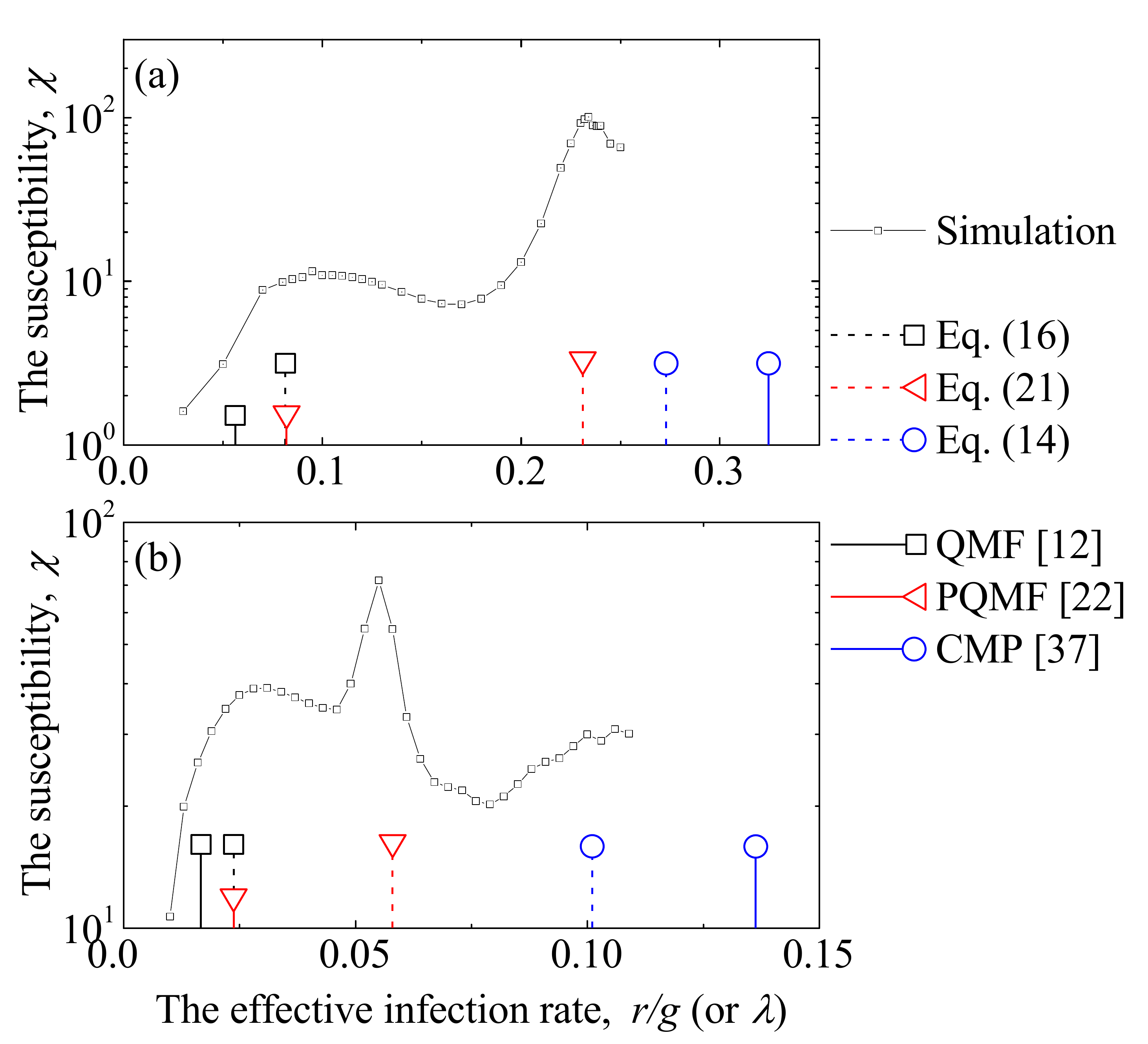}
\caption{The susceptibility $\chi$ as a function of the effective infection rate $\lambda$ on single realizations of a scale-free network. Parameters: minimum degree $3$; (a) $\gamma=4.5$, $N=10^7$, $k_{\max}=314$, and $\langle l\rangle=6.841$; (b) $\gamma=3.5$, $N=3\times10^7$, $k_{\max}=3611$, and $\langle l\rangle=2.789$. The network of (a) is the same as that in Fig.~\ref{fig.2}. To get smooth simulation curves, the relaxation time $t_r$ are varied from $10^5$ to $10^9$ depending on $N$ and $\lambda$. We show several analytical estimates of the transition point.
}\label{fig.6}
\end{figure}

We return the epidemic threshold prediction on a scale-free network with $\gamma>3$, and summarize the results in Fig.~\ref{fig.6}. 
First, we can predict the global peak by using the Eq.~\eqref{e.14} with the global coupling $x\to0$, and show that our results are less than the prediction of~\cite{PhysRevX.10.011070} where the latter tends to zero in the thermodynamic limit.
Second, we can predict the first localized peak by using the Eq.~\eqref{e.16} and $k=k_{\max}$. Note that they almost coincide between our results (black dashes) and the prediction of PQMF (red lines). Third, we can predict the second localized peak by using the Eq.~\eqref{e.21} whose parameters are set as $k_1=k_{\max}$, $k_2\sim k_{\max}$, and $l=\langle l_i\rangle$. Compared with $l$, the value of $k_2$ has less influence on $f(k_1,k_2,r,g)$ in Eq.~\eqref{e.20}, which is why we use an approximation $k_2\sim k_{\max}$ in here.
The $l_i$ is the distance between a star subnetwork with center node $i$ and the $k_{\max}$-star subnetwork.
We average the $l_i$ of nodes whose degree is greater than $k'$, where $k'$ satisfy $\sum_{k=k'}^{k_{\max}}kP(k)N\sim0.001N$.
This value of $0.001N$ is selected to ensure that the number of continuously activated subnetworks is sufficient to generate fluctuations of prevalence, but not reach the global spreading.
Although the prediction we gave has many rough approximations, it is able to roughly hold the position of the second localized peak in Fig.~\ref{fig.6}.

Finally, we would like to point out that the nodes with a small degree still cannot be continuously active in Fig.~\ref{fig.2}(b), though its infection rate is higher than the second peak of Fig.~\ref{fig.6}(a).

\section{Conclusions}
In conclusion, we have proposed a high-accuracy theoretical approach for analyzing the SIS model on networks that takes both network structure and dynamic correlations into account. Our approach works from population-wide outbreaks, over localized, to threshold. For population-wide outbreaks, we predict the average risk of infection for neighbors of nodes with degree $k$ in Fig.~\ref{fig.2} and predict more accurate prevalence in Fig.~\ref{figrho}. For localized spreading, we give a high-accuracy prediction for the 
multiple peaks and localized prevalence on the two star network. For epidemic threshold, we predict the multiple localizd peaks and global threshold on scale-free networks. Below the global threshold, hubs can keep the epidemics alive for an extended period of time. This phenomenon is known from medical epidemiology as well~\cite{yorke1978dynamics}, and the explanation for why some diseases be endemic even though they are under the threshold.

%In addition to predict the multiple peaks in susceptibility, our methods improves existing theories when it comes to calculating the prevalence close to the transition in heterogeneous networks. Our method can be easily be extended to other spreading dynamics on heterogeneous networks. 
%In the Supplemental material, we show initial simulation results for threshold~\cite{PhysRevE.98.012310} and contact~\cite{PhysRevE.84.066102} processes.

\acknowledgments
This work was supported by the National Natural Science Foundation of China (Grant No.\ 11705147 and No.\ 11975111). P.H. was supported by JSPS KAKENHI Grant Number JP 18H01655.

\appendix
\makeatletter
\@addtoreset{equation}{section}
\@addtoreset{figure}{section}
\makeatother
\renewcommand{\theequation}{A\arabic{equation}}
\renewcommand{\thefigure}{A\arabic{figure}}
\section{}\label{App:A}
\begin{figure}
\includegraphics[width=\linewidth]{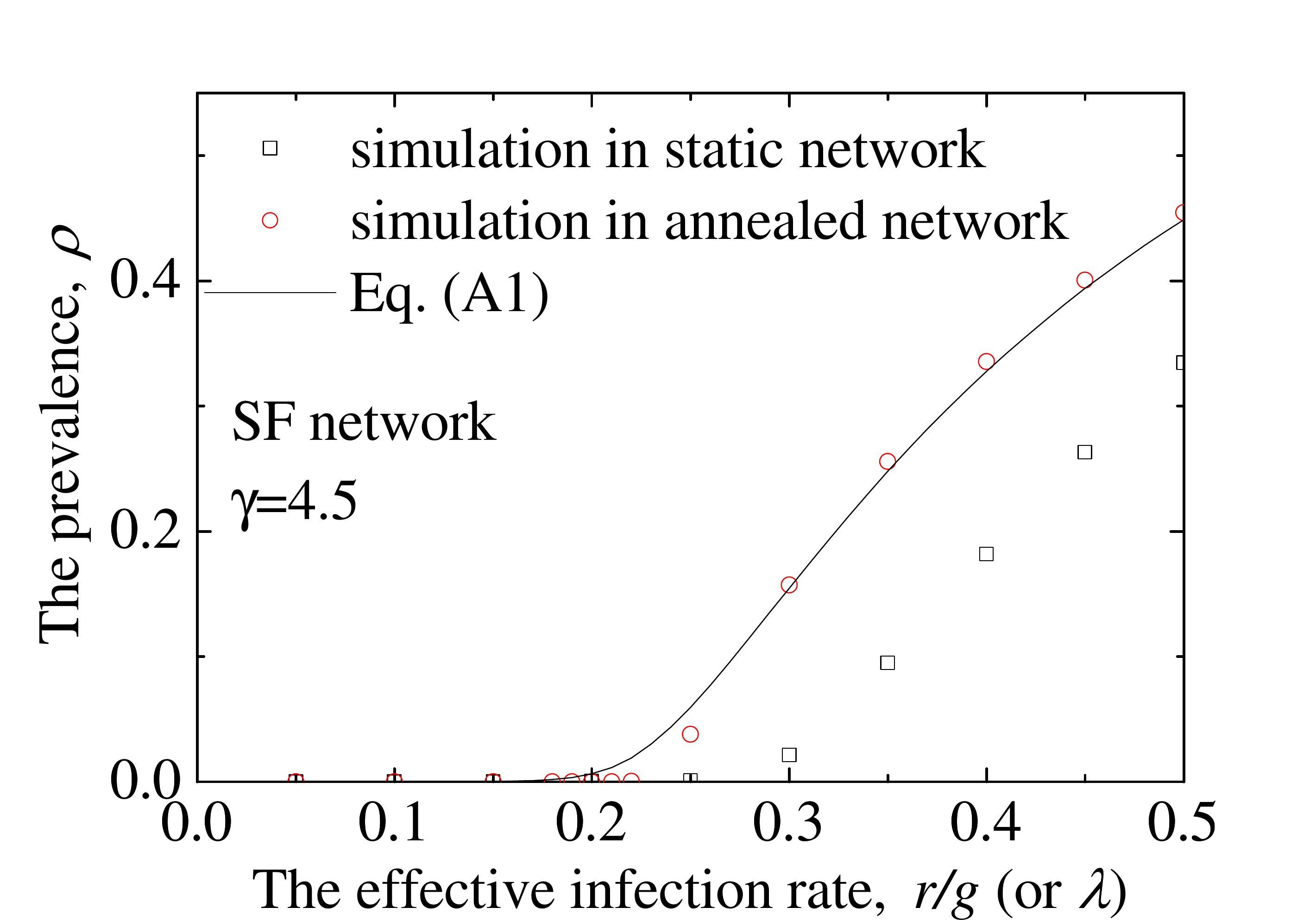}
\caption{(Color online) The epidemic prevalence $\rho$ in the SIS process is plotted as a function of the effective infection rate on annealed and static scale-free networks. Parameters: $N=10^5$, minimum degree $3$ and $\gamma=4.5$. Lines are from Eq.~\eqref{eq.a1}, while the points are simulation results.}\label{fig.A}
\end{figure}

The quenched mean-field theory is an individual-based mean-field theory. It takes into account the network structure fully but ignores the dynamic correlation. We denote the infected probability of an individual $i$ by $I_i$, and the QMF dynamic equations can be given by
\begin{equation}
\label{eq.a1}
\frac{dI_i}{dt}=-gI_i+r S_i\sum_j I_j A_{ij},
\end{equation}
where $A_{ij}$ is the adjacency matrix with value $A_{ij}=1$ if individuals $i$ and $j$ are connected, and zero otherwise. Here, we can obtain the epidemic prevalence $\rho$ by iterating the large set of Eq.~\eqref{eq.a1} to the stable state, which is limited by the large network size. From Fig.~\ref{fig.A}, we can see that the $\rho$ obtained from QMF agrees well with the simulation results from annealed networks instead of those from static networks.

\renewcommand{\theequation}{B\arabic{equation}}
\renewcommand{\thefigure}{B\arabic{figure}}
\section{}\label{App:B}
%\subsection{The discrete probability distributions on star subnetwork}
\begin{figure}
\includegraphics[width=\linewidth]{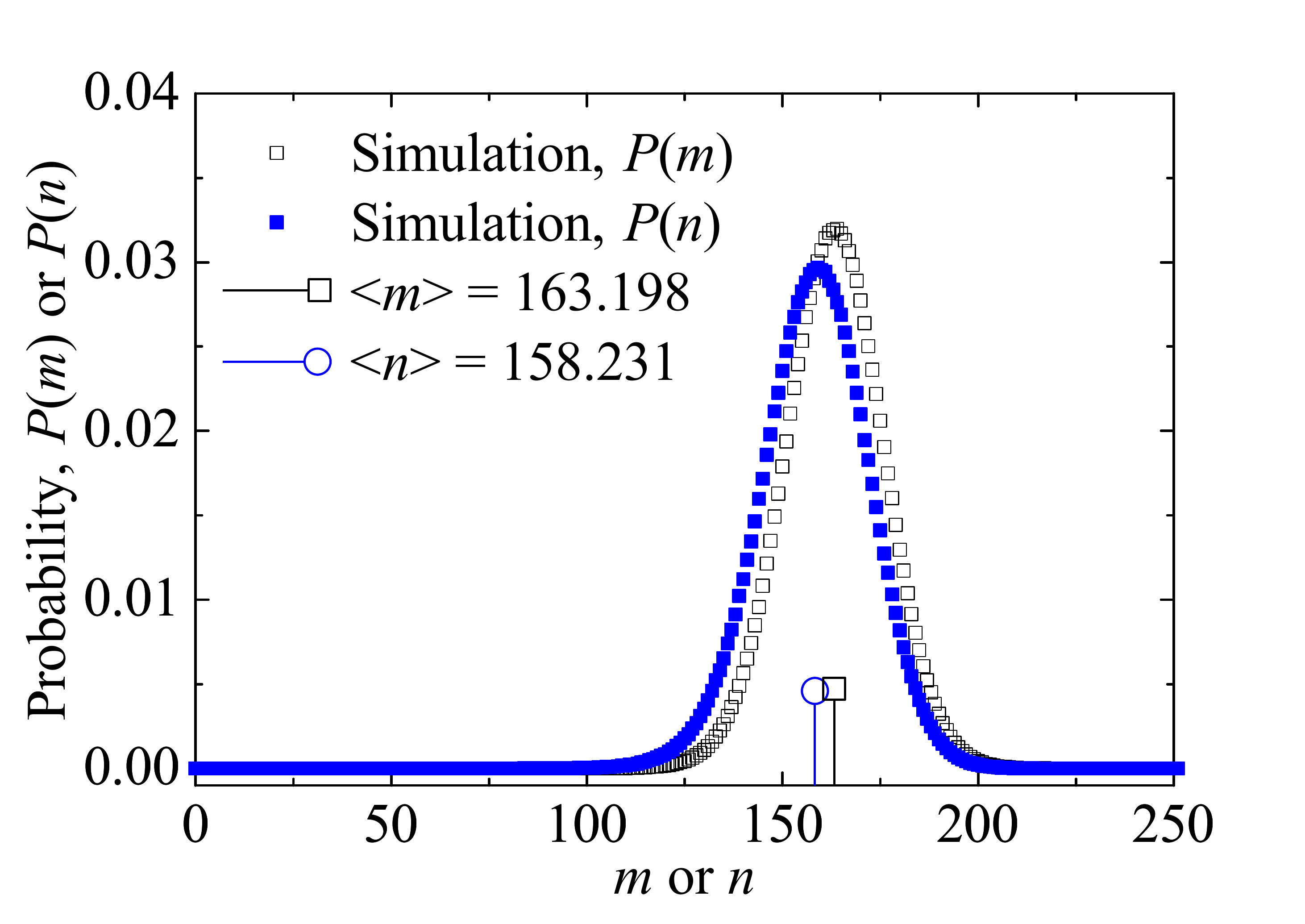}
\caption{(Color online) Epidemic spreading on a star network with one permanently infected
leaf node. The probabilities $P(m)$ and $P(n)$ is plotted as a function of the number of infected leaf nodes when the center node is cured and is infected. Parameters: $r=0.2$, $g=1$, network size $N=1000$.}\label{fig.B}
\end{figure}

The $P(m)$ is denoted as the probability distribution function of the number of infected leaf nodes $m$ when the central node is cured, while the $P(n)$ is denoted as the probability distribution function of the number of infected leaf nodes $n$ when the central node is infected.
In Fig.~\ref{fig.B}, we record $10^7$ samples continuously for the two variables $m$ and $n$ after a relaxation time $t=1000$ on the star network with one permanently infected leaf node.
From Fig.~\ref{fig.B}, we can see that the $P(m)$ and $P(n)$ are both peak distributions. For the sake of simplicity, we use these averages $\langle m\rangle=\sum_m mP(m)$ and $\langle n\rangle=\sum_n nP(n)$ as approximate to replace the $P(m)$ and $P(n)$. In the main body, $\langle m\rangle$ and $\langle n\rangle$ are writen as $m$ and $n$, respectively. 

\renewcommand{\theequation}{C\arabic{equation}}
\renewcommand{\thefigure}{C\arabic{figure}}
\section{}\label{App:C}

For random regular networks, the epidemic threshold gives $\lambda_c=1/(k_0-1)$ which can be found by using \textit{pair approximation}, \textit{dynamic correlation}, or \textit{heterogeneous pair-approximation}. And then, we can calculate
\begin{equation}\label{eq.c1}
\frac{\lambda_{c_1}(\text{$k_{\max}$})}{\lambda_c}\sim(k_0-1)\sqrt{\frac{2}{k_0}}>1
\end{equation}
for all degrees $k_0>2$.

For Erd\H{o}s-R\'enyi networks, the epidemic threshold is $1/\langle k\rangle$ which can be found by using \textit{dynamic correlation} or \textit{heterogeneous pair-approximation}. And then, we can calculate
\begin{equation}\label{eq.c2}
\frac{\lambda_{c_1}(\text{$k_{\max}$})}{\lambda_c}\sim\langle k\rangle\sqrt{\frac{2}{k_{\max}}}.
\end{equation}
Combining $\langle k\rangle\sqrt{2/k_{\max}}>1$ with degree distribution $P(k)=\frac{\langle k\rangle^k}{k!}\exp^{-\langle k\rangle}$ and the cut-off condition of homogeneous networks $P(k_{\max})N=1$, we can obtain the condition 
\begin{equation}\label{eq.c3}
N<\exp^{\langle k\rangle}\frac{(2\langle k\rangle^2)!}{\langle k\rangle^{2\langle k\rangle^2}}
\end{equation}
that satisfy relation $\lambda_c<\lambda_{c_1}(\text{$k_{\max}$})$. For example, for the ER networks with $\langle k\rangle=4$ and $6$, respectively, the population size $N<7\times10^{17}$ and $N<2\times10^{50}$ can satisfy relation $\lambda_c<\lambda_{c_1}(\text{$k_{\max}$})$.

For a large size scale-free networks with $2<\gamma<2.5$, we can easily obtain the relations of $k_{\min}^{-\gamma+2}\gg k_{\max}^{-\gamma+2}$ and $k_{\max}^{-\gamma+3}\gg k_{\min}^{-\gamma+3}\geq k_{\min}^{-\gamma+2}$ when $k_{\max}\gg k_{\min}$. And then, we can calculate
\begin{equation}\label{eq.c4}
\lambda_c^{\mathrm{DC}}=\frac{\langle k\rangle}{\langle k^2\rangle-\langle k\rangle}\sim k_{\max}^{\gamma-3}<k_{\max}^{-0.5}.
\end{equation}
Similarly, we can also calculate
\begin{equation}\label{eq.c5}
\lambda_c^{\mathrm{QMF}}=\frac{1}{\Lambda}\sim\frac{\langle k\rangle}{\langle k^2\rangle}\sim k_{\max}^{\gamma-3}<k_{\max}^{-0.5},
\end{equation}
where $\Lambda$ is the largest eigenvalue of the adjacency matrix.
Combining with the relation $\lambda_{c_1}(\text{$k_{\max}$})\sim\sqrt{2}k_{\max}^{-0.5}$, we can obtain the relations $\lambda_c^{\mathrm{DC}}<\lambda_{c_1}(\text{$k_{\max}$})$ and $\lambda_c^{\mathrm{QMF}}<\lambda_{c_1}(\text{$k_{\max}$})$ on the SF network with $2<\gamma<2.5$.

\bibstyle{abbrv}
\bibliography{bbb}

\end{document}